\documentclass[journal]{IEEEtran}

\usepackage[utf8]{inputenc}
\usepackage[english]{babel}

\usepackage{cite}
\usepackage{amsmath}
\usepackage{mathrsfs}
\usepackage{array}

\usepackage{subfig}

\usepackage{graphicx}
\usepackage{enumitem}
\usepackage{cite}
\usepackage{amsthm}
\usepackage{amssymb}

\newtheorem{theorem}{Theorem}

\newtheorem{lemma}{Lemma}

\usepackage{lipsum}
\usepackage{multicol,blindtext}
\usepackage{mathtools, cuted}
\usepackage{lipsum, color}
\usepackage{algorithm}
\usepackage[noend]{algpseudocode}
\usepackage{comment}

\makeatletter
\def\BState{\State\hskip-\ALG@thistlm}
\makeatother

\newcommand{\clN}{{\cal N}}

\usepackage{url}

\begin{document}

\title{Voltage Inference for and Coordination of Distributed Voltage Controls in Extremely-High DER-Penetration Distribution Networks}

\author{Ying Xu,~\IEEEmembership{Member,~IEEE,}
		and Zhihua Qu,~\IEEEmembership{Fellow,~IEEE,}
}
\maketitle

\IEEEpeerreviewmaketitle
\begin{abstract}
   The unique problems and phenomena in the distributed voltage control of large-scale power distribution systems with extremely-high DER-penetration are targeted in this paper. 
    First, a DER-explicit distribution network model and voltage sensitivity are derived. Based on that, a voltage inference method is implemented to fill the gap of measurement insufficiency in the grid-edge areas. Then, autonomous Q control being implemented in each local area, a $\overline{Q}$-coordinated P control is designed to coordinate the reactive and real power controls. 
    All the algorithms have been tested in standard and synthetic systems, and have expected results. 
    Moreover, an open-source software platform, which integrates the modeling of communication networks, DER controls, and the power networks, is developed to enable the distributed control and optimization algorithms in the grid simulation of the large-scale distribution systems.

\end{abstract}

\section{Introduction}

The environmental, economic, and social sustainable challenges for human beings have become more intensive than ever before, which forces the fossil-energy-dominated power industry to adopt and develop advanced technologies for integrating diverse generation, energy-efficiency, and clean energy resources.

Without the natural and geographic limitations of hydro-power, the small-size, distributed energy resources (DERs) such as solar and wind are preferable and available to energize communities and businesses around the world.
DER-based renewable power generation has been seen growing at a surprisingly fast pace.
It is believed by many that the renewable generation will gradually take the share of traditional electricity generation (mainly by large-size synchronous generators) and will eventually dominate the power grid in the future.

In the context of voltage control of a large-scale distribution system with extremely-high renewable penetration, many challenges rise as the percentage of DER generation is continuously increasing.
The existing power systems are mainly operated on centralized control (i.e., SCADA/EMS/DMS, short for Supervisory Control And Data Acquisition/Energy Management Systems/Distribution Management Systems) which focuses on a limited number of major generators and large controllable devices, while the smaller loads or devices at grid-edge are not taken into account.
In the on-going development of power grid infrastructure, more power electronics, measurements, and communications will be introduced into the grid edge, which will offer more flexibility to identify, control, and manage the power flow and electricity quality of the future grid. While the current control centers and their support ICT systems are yet able to handle this situation: measuring, communicating, controlling and managing thousands to millions of active devices,
a set of operational tools are necessary to better integrate and control distributed resources and to maintain the robustness and resilience of power grids.

With the capability to control million device systems, distributed methods attracted significant research interests\cite{Alyami14,Plytaria17,Chai18}. The distributed algorithms in power systems have the following advantages: they have faster solving speed through parallel computing which decomposes large-scale system (or high-order) problems into smaller ones; they require simpler communication structure and less information transportation; they are more resilient due to the self-organized design which is free of the centralized controller; the agent-based and modular design can better protect users' privacy.
In the past decades, tremendous research efforts have been made to implement distributed methods into power systems, as reviewed in \cite{molzahn2017survey}. In \cite{xin2011cooperative,maknouninejad2014realizing,xin2011self,Xia19}, distributed methods are implemented into smart grids and reported to achieve better performance.

However, there are many obstacles from theory to practice for these distributed algorithms to be implemented in real-world systems. First, insufficient measurement in the grid-edge areas of a large-scale system can challenge most of the existing distributed algorithms.
Second, the unbalance between the power sources and loads can be huge when the system scale is large, which will cause substantial voltage violations on both the high- and low- limit sides. Moreover, the strategy of the coordination between distributed Q and P control and its feasibility and scalability need to be thoroughly studied as well.
Motivated by all the above issues, a two-level, fully distributed control design is developed in this paper.
In this design, a voltage inference method is first derived for the observability increase in the grid-edge areas, which are traditionally weakly managed (in terms of communication and control) in the existing framework of power control systems. Then a self-organized and autonomous Q control is implemented in each local area, and a $\overline{Q}$-coordinated P control is designed based on the critical system-level information, which is obtained using the cooperative maximum and minimum protocols.
The effectiveness and advantage of the method have been tested on different large-scale systems, including IEEE 8500-node and synthetic 100k-node systems.
The main contributions of this paper are briefly summarized as follows:
\begin{itemize}
    \item A DER-explicit distribution network model and the network voltage sensitivity are derived, which form the foundations of large-scale distribution network control;
    \item A voltage inference method is developed for the control to cover the grid-edge areas;
    \item The autonomous Q control is developed and implemented in each local area;
    \item A distributed $\overline{Q}$-coordinated P control is designed to minimize the real power curtailment;
    \item An open-source software platform is developed to enable the distributed control and optimization algorithms in the grid simulation of the large-scale distribution systems.
\end{itemize}

The rest of the paper is organized as follows: Section II is the problem formulation; section III provides a DER-explicit distribution network model and the voltage inference method based upon the model; the control design for the large-scale systems with extremely high penetration of renewables is derived in section IV; case study and simulation results are shown in section V; at last, section VI concludes the paper.

\section{Problem Formulation and Preliminaries}

\begin{figure}[h!]
        \centering
        \includegraphics[width=0.8\linewidth]{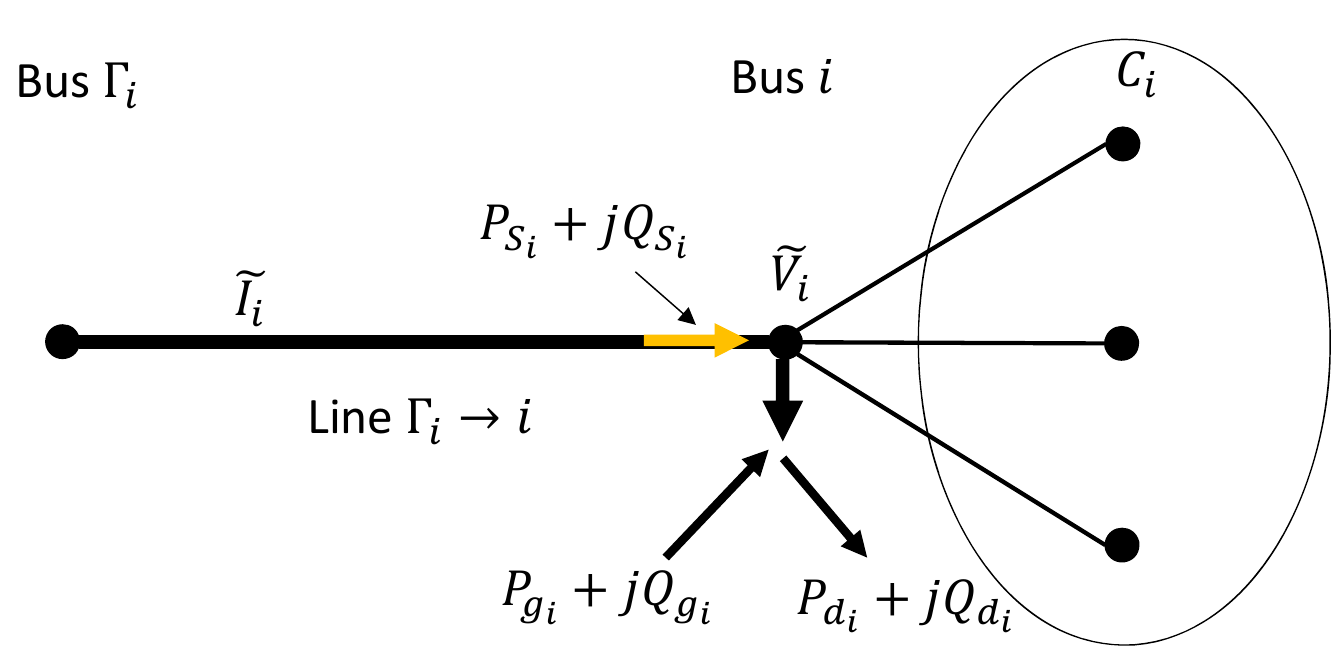}
        \caption{A segment of a distribution network}
        \label{fig:busI}
    \end{figure}

    Consider a radial distribution power system shown in figure \ref{fig:busI}, where $\tilde{V}_i$ is the AC voltage at bus $i$. Symbol $\Gamma_i$ denotes the parent bus of bus $i$, ${\mathcal C}_i$ represents the set of its children buses, $R_i+jX_i$ is the impedance of line segment $\Gamma_i \rightarrow i$, $\tilde{I}_i$ is the AC current flowing over the line from bus $\Gamma_i$ to bus $i$,
    and $(P_{s_i}+jQ_{s_i})$ is the complex power supplied to bus $i$ through line segment $\Gamma_i \rightarrow i$.

    The net power injection at bus $i$  is represented by
    \begin{equation}
    P_i + j Q_i = -(P_{g_i} + j Q_{g_i}) +(P_{d_i} + j Q_{d_i}),\label{eq:businj}
    \end{equation}
    where $(P_{g_i} + j Q_{g_i})$ and $(P_{d_i} + j Q_{d_i})$ are the DER injection and load, respectively.
    As in \cite{xin2011self}, inverter-based DERs are typically fast enough to regulate its active and reactive power outputs and, by using phase locked loops (PLL) and D-Q decoupling inverter controls,
    the active/reactive power injections $P_{g_i}$ and $Q_{g_i}$ can be viewed as the equivalent control inputs for distribution networks. In this paper, these controls will be designed within the following operational limits while meeting all the control objectives:
    \begin{equation}
        0\leq P_{g_i} \leq \overline{P}_{g_i},\;\;\;
        |Q_{g_i}| \leq \overline{Q}_{g_i}\equiv
         \sqrt{\overline{S}_{i}^2-P_{g_i}^2}, \label{pwr_cs}
    \end{equation}
    where $\overline{P}_{g_i}$ is the maxim available real power output at the $i$th DER, and $\overline{S}_{i}$ is the apparent power capacity of its inverter.

    The voltage stability of power systems is generally described as
    \begin{equation}
       V_i \in [\underline{V}, \overline{V}],\label{eq:reg}
    \end{equation}
    where $\underline{V}$ and $\overline{V}$ are the operational limits. Their typically values are  $\underline{V}=0.95$ and  $\overline{V}=1.05$. There are two methodologies to address the voltage regulation problem. One is to include \eqref{eq:reg} into the optimal power flow (OPF) problem and solve it distributively \cite{liu2016distributed,dall2016optimal,dall2015photovoltaic,jabr2017linear}. The other is to satisfy \eqref{eq:reg} directly by implementing a distributed cooperative voltage control \cite{ecc2020}.

    As the penetration level is raised from low (say, the percentage in teens) to high (e.g., over $50\%$ or even $100\%$), the spatial profile of network voltages becomes highly nonlinear (rather than a gradual drop trend from the feeder to the grid edge), has a much larger range, and is dynamic. This brings about several operational challenges arise, and the following are addressed in the paper.

\begin{figure}[h!]
        \centering
        \includegraphics[width=0.8\linewidth]{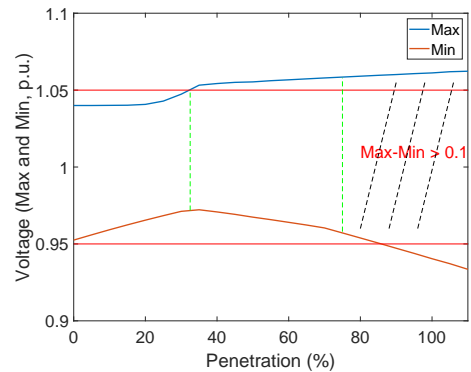}
        \caption{Max/min voltages of IEEE 8500 system}
        \label{fig:vv}
    \end{figure}

\begin{figure}[h!]
        \centering
        \includegraphics[width=0.75 \linewidth]{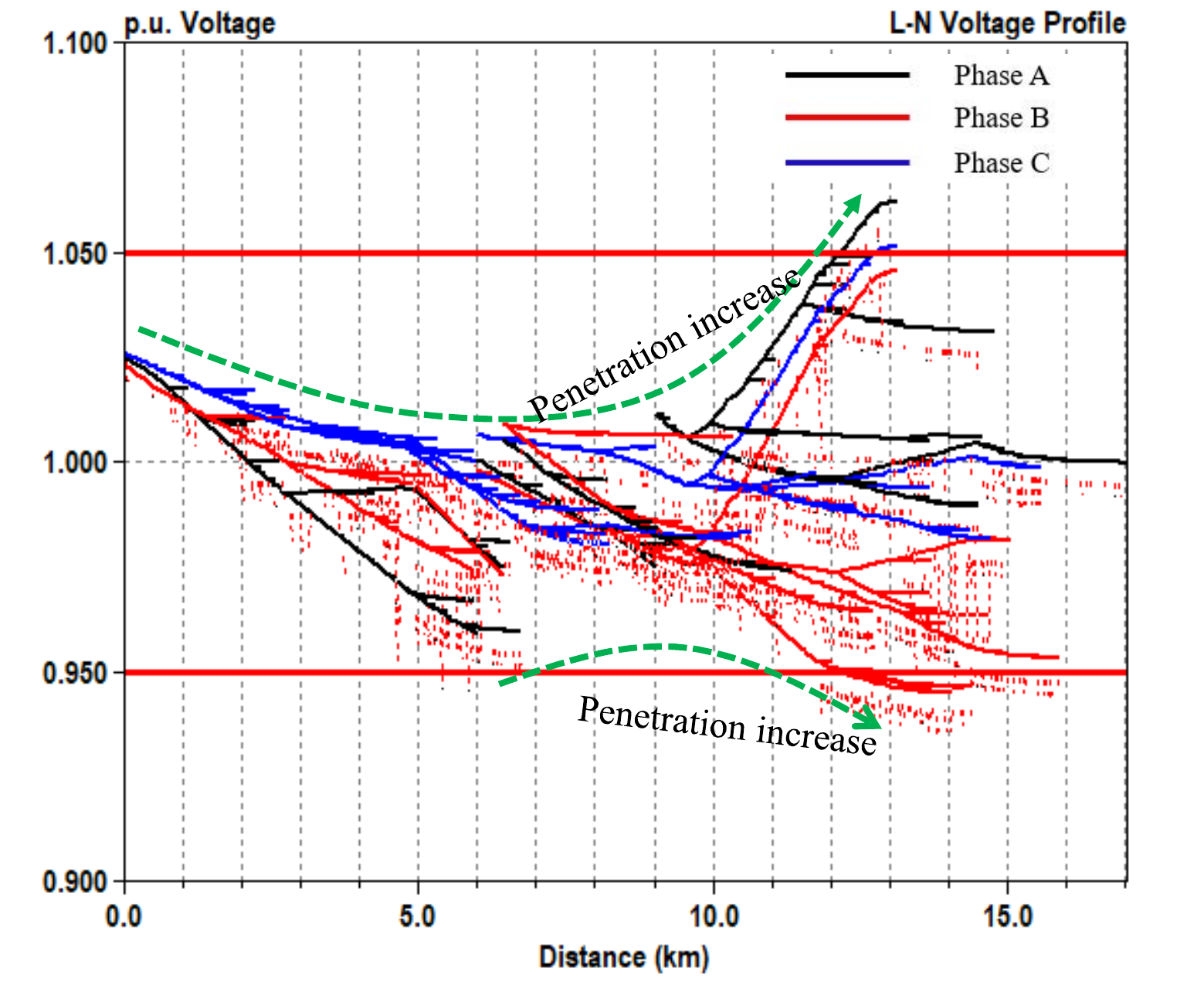}
        \caption{Locations of max/min voltages of IEEE 8500 system -- migration trajectories (in green) with respect to penetration level.}
        \label{fig:vv-location}
    \end{figure}

    {\it 1) Variant Voltage Profile under High Penetration}: The voltage profile is primarily due to the spatial imbalance between power injections and loads. In traditional distribution networks, the voltage profile has a gradual downward trend, and the low voltage problem can be corrected by a few step-up transformers along the radial lines. As penetration goes up, the  maximum and minimum voltages in the network may become worse simultaneously, and their locations may be separated by a large number of nodes. As $Q$ controls attempt to suppress higher voltages in the network, they may also aggregate lower voltages elsewhere in the network. As an example, consider the case of the IEEE 8500 system with high-penetration of DERs \cite{ecc2020}.
    Figure \ref{fig:vv} shows the trends of maximum and minimum voltages with respect to the penetration levels(with limited amount of reactive power compensation), and figure \ref{fig:vv-location} shows the corresponding movements of the maximum and minimum voltage locations. It is clear from the two figures that      
    the voltage problem arises once the penetration exceeds certain level (the first green vertical line) and that the locations of voltage violations may be changing.

    {\it 2) Information Acquisition and Sharing about the Network's Voltage Profile}: Due to intermittency of renewable generations, the aforementioned profile will be dynamically changing, hence it is necessary to monitor critical voltages real-time. This task is complicated by the facts that, in most distribution networks, voltages at most of the nodes are not measured real time and that local communication can at best be expected. Hence, it is crucial to develop a way for every local control to learn maximum and minimum voltages, possibly occurring at the nodes that are not continuously measured.

    {\it 3) Limitations of Distributed $Q$ Controls}: As shown in \cite{xin2011self}, reactive power controls at neighboring DERs can be pooled together to either suppress the maximum voltage or elevate the minimum voltage. Nonetheless, the consensus-based control by itself cannot do both, and all reactive power controls have limited capacities. A more comprehensive strategy is needed to dynamically narrow voltage deviations into allowable operational range while maximizing the active power injections.

    The proposed design aims to address the aforementioned changes by developing and incorporating the following two key innovations.


    {\it a) Local Acquisition and Sharing about Max/Min Voltages by Either Measurement or Inference}: A distributed communication and control architecture may involve overlaping clusters. Within each cluster, DERs can communicate with their neighbors and share information. It is shown in \cite{qu2009cooperative} that, as long as the overall communication topology is strongly connected, the knowledge of maximum and minimum voltages in the network can be gained through local communication by employing the distributed maximum and minimum protocols. Since many nodes in distribution networks have no real-time measurement, a systematic approach of inference needs to be developed to determine the highest/lowest voltages, which is the subject of section \ref{sbsct:sa}.

    {\it b) Voltage Stability through Local Coordination of  both $Q$-Control and $P$-Control}: The proposed voltage
    control is built upon the distributed subgradient-based cooperative $Q$ control, and this $Q$-$V$ control is a fast-responding local control, working within each of the clusters. The proposed control has another distributed layer of coordination between $P_{g_j}$ and $Q_{g_i}$. This new layer takes advantage of the acquired information on max/min voltages within the whole distribution network and autonomously adjusts active power injections across multiple clusters in the event that either reactive power injections are exhausted already or their effects cannot move both the  highest and lowest voltages in the right directions.
    The designs of these distributed controls are the topic of section \ref{sec:ctrl}.

\section{Voltage Inference Based on Sensitivity Analysis}\label{sbsct:sa}

    In order to address the main challenges identified in the previous section, we begin with developing a DER-explicit distribution network model. It follows from Appendix \ref{appen:model} that voltage and power injection equations of the distribution network are: for any $k\in \clN_b\stackrel{\triangle}{=} \{1,\cdots,n_b\}$ with $n_b$ being the total number of buses,
    \begin{subequations}
    \begin{align}
    V_k^2 = & \; \sum_{i\in \clN_b}(2\mathcal{R}_{ki}P_{g_i}+ 2\mathcal{X}_{ki}Q_{g_i}
        - 2\mathcal{Z}'_{ki} P_{L_i})+V_0^2 \nonumber \\
        & \; -\sum_{i\in \clN_b}(2\mathcal{R}_{ki}P_{d_i}+ 2\mathcal{X}_{ki}Q_{d_i}), \label{eq:vpl}\\
     P_{L_k} = & \;
     \frac{R_k}{V^2_k}\left[\left(P_{g_k}+P_{d_k}+\sum_{m\in \mathcal{D}_k}
            (P_{g_m}+P_{d_m}+P_{L_m})\right)^2\right.\nonumber\\
        & 
        \left.+\left(Q_{g_k}+Q_{d_k}+\sum_{m\in \mathcal{D}_k}(Q_{g_m}+Q_{d_m}+b_m P_{L_m})\right)^2 \right]\label{eq:plv},
    \end{align}\label{eq:brchflowvpl}\noindent
    \end{subequations}
    where $V_0$ is the voltage at feeder head; $V_k$ is the voltage at bus $k$; $V_i$ is bus voltage at bus $i$; $P_{L_k}$ is the loss on line segment $\Gamma_k\to k$;
    ${\mathcal P}_{0k}$ denotes the unique path from the feeder (bus $0$) to bus $k$, and $\mathcal {P}_{0j}\cap \mathcal {P}_{0k}$ is the common path segments from bus $0$ to both buses $j$ and $k$; $\mathcal{D}_k$ is the bus set of the subtree after bus $k$; and  $\mathcal{R}_{ki}$, $\mathcal{X}_{ki}$, and $Z'_{ki}$ are constant network parameters defined as follows:
    \begin{equation*}
    \begin{split}
        &\mathcal{R}_{ki}=\sum_{l\in \mathcal{P}_{0i}\cap\mathcal{P}_{0k}}R_l,\;\;\mathcal{X}_{ki}=\sum_{l\in \mathcal{P}_{0i}\cap\mathcal{P}_{0k}}X_l,   \\
        & \mathcal{Z}'_{ki} =\mathcal{R}'_{ki}+b_i\mathcal{X}'_{ki},\;\;b_i=X_i/R_i,\\
        &\mathcal{R}'_{ki}=
        \begin{cases}
            \mathcal{R}_{ki} - R_i, &\text{ if } k\in {\cal D}_i\\
            \mathcal{R}_{ki} &\text{ otherwise}
        \end{cases},\;\;
            \\
        &\mathcal{X}'_{ki}=
        \begin{cases}
            \mathcal{X}_{ki} - X_i, &\text{ if } k\in {\cal D}_i\\
            \mathcal{X}_{ki}, &\text{ otherwise}
        \end{cases}.
    \end{split}
    \end{equation*}
    Like the branch model, the above voltage and power flow model in \eqref{eq:brchflowvpl} is also branch oriented and has no angle variables involved. Moreover, it has several distinctive features: i) The model is both exact and explicit in terms of all DER injections $P_{g_i}$ and $Q_{g_i}$; ii) The nonlinear model describes how an injection anywhere in the radial network will affect bus voltages as well as losses at different line segments; iii) The model is scalable with respect to both the numbers of buses and DERs; iv) The model parameters can be assumed as known for at DER buses, and they represent the invariance embedded in the distribution network; v) While the equations are nonlinear, they need not to be solved real-time; instead, their sensitivity analysis yields useful equations for voltage inference, which is the subject of the following lemmas.

    \begin{lemma}\label{lmvpq1} For any radial distribution network,
    \begin{align}
     \frac{\partial V_k}{\partial P_{g_j}} =&\frac{\mathcal{R}_{kj}}{V_k}-  \frac{1}{V_k}\sum_{i\in \clN_b}\mathcal{Z}'_{ki}\frac{\partial P_{L_i}}{\partial P_{g_j}} \label{dvdp}\\
     \frac{\partial V_k}{\partial Q_{g_j}}  =&\frac{\mathcal{X}_{kj}}{V_k}-  \frac{1}{V_k}\sum_{i\in \clN_b}\mathcal{Z}'_{ki}\frac{\partial P_{L_i}}{\partial Q_{g_j}}\label{dvdq} \end{align}
    where
    \begin{align}
        \frac{\partial P_{L_i}}{\partial P_{g_j}} = &- \frac{1}{V_i} P_{L_i}\frac{\partial V_i}{\partial P_{g_j}} \nonumber\\
     &-  \frac{R_i}{V^2_i} \left(P_{s_i}+\sum_{m \in {\cal T}_i} (P_{s_i}+Q_{s_i}b_m)\frac{\partial P_{L_m}}{\partial P_{g_j}}\right),   \\
      \frac{\partial P_{L_i}}{\partial Q_{g_j}} = &- \frac{1}{V_i} P_{L_i}\frac{\partial V_i}{\partial Q_{g_j}} \nonumber\\
     &-  \frac{R_i}{V^2_i} \left(Q_{s_i}+\sum_{m \in {\cal T}_i} (P_{s_i}+Q_{s_i}b_m)\frac{\partial P_{L_m}}{\partial Q_{g_j}}\right).
    \end{align}
    \end{lemma}

    \begin{lemma}\label{lmvpq2} Under the assumptions that the losses are ignored and that bus voltages are all close to $1$ p.u., then
    \begin{equation}
     \frac{\partial V_k}{\partial P_{g_i}}\approx \mathcal{R}_{ki}, \;\;\; \text{and} \;\;\;  \frac{\partial V_k}{\partial Q_{g_i}}\approx \mathcal{X}_{ki}, \label{eq:zetaxi}
    \end{equation}
    or equivalently,
    \begin{equation}
        \Delta V_k=\sum_{i\in \clN_b}(\mathcal{R}_{ki}\Delta P_{g_i}+ \mathcal{X}_{ki}\Delta Q_{g_i}).\label{eqvklnr}
    \end{equation}
    \end{lemma}

\begin{figure}[h!]
        \centering
        \includegraphics[width=\linewidth]{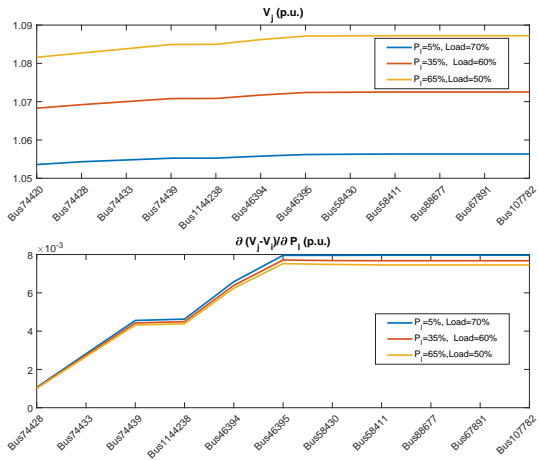}
        \caption{Bus voltages and the sensitivities with respect to different loading and penetration levels of real power injections at bus $i$($\#74428$) and bus $l$($\#46395$) in EPRI Ckt5 \cite{montenegro2015multilevel}.}
        \label{ckt5}
    \end{figure}

    Lemma 1 is a direct result of differentiating equation \eqref{eq:vpl} in which the loss sensitivities are defined and can be calculated in a recursive form. Without a wide-area monitoring system (of expansive measurement and communication capabilities), these loss sensitivities are difficult to evaluate. To reduce the information needed, lemma 2 provides an approximate but more practically useful result. For distribution networks operating under standard conditions, the approximation in lemma 2 is reasonable, as illustrated by the top plot in figure \ref{ckt5}. Although the node voltages themselves and the differences of neighboring nodal voltages are changing as {the load and penetration levels change},  the sensitivity values of neighboring voltage differences with respect to the power injection are invariant for all the nodes between two DER locations, as shown in the bottom plot of figure \ref{ckt5}. This inherent invariance makes it possible to perform voltage inference. 
    
    To be explicit about voltage inference, consider a distributed generation of injections $P_{g_k}$ and $Q_{g_k}$. With the DERs at nodes $i$ and $l$ and in the same cluster, their voltages $V_i$ and $V_l$ are measured real time. Suppose that node $j$ is also in the same cluster and along the same radial line as nodes $i$ and $l$ but its voltage is not measured real time. With standard digital meters, the load and voltage at the $j$th node are monitored periodically, say, every 15 or 20 or 30 minutes at time instants $t_k$. These recent data make it possible for DERs in the same cluster to infer its current voltage value. Specifically, it follows that, at node $i$, we know that, for $t\in[t_k, \; t_{k+1})$,
    \begin{equation}
        V_j(t) \approx  V_i(t) + \frac{V_l(t)-V_i(t_k)}{V_l(t_k)-V_i(t_{k})}[V_j(t_k)-V_i(t_k)]
        \stackrel{\triangle}{=}  \hat{V}_j(t),
        \label{eq:delta_V}
    \end{equation}
   
   Estimation $\hat{V}_j(t)$ gives us an instantaneous prediction of the incremental change of any node voltage (if not measured), and it is periodically corrected.


  Sensitivity expression \eqref{eqvklnr} also reveals other interesting observations:
    \begin{itemize}
    \item[i)] The voltage variations induced by real power injections in the network can be compensated by reactive power injections up to their capacities. Coordinated $Q$ controls, such as distributed subgradient-based cooperative control \cite{maknouninejad2014realizing}, are more appropriate.
    \item[ii)] Once the reactive power capacities are reached, voltage regulation has to be done through real-power curtailments which also  increase reactive power capacity. That is, coordination between $Q$ controls and $P$ controls are needed, as voltages exceed both the upper and lower bounds or their range becomes large than $0.1$ pu.
    \item[iii)] Each DER can deduct the current voltages at its neighboring and non-injection nodes by using their nominal values and its own voltage variation.
    \item[iv)] Instead of measuring voltages at many nodes and communicating the information across the network, it is only necessary to determine the minimum and maximum voltages in the network. These limited information can easily be spread and shared through local communication networks.
    \end{itemize}

\section{Synthesis of Coordinated $Q$ and $P$ Controls} \label{sec:ctrl}
\begin{figure}[h!]
        \centering
        \includegraphics[width=0.75 \linewidth]{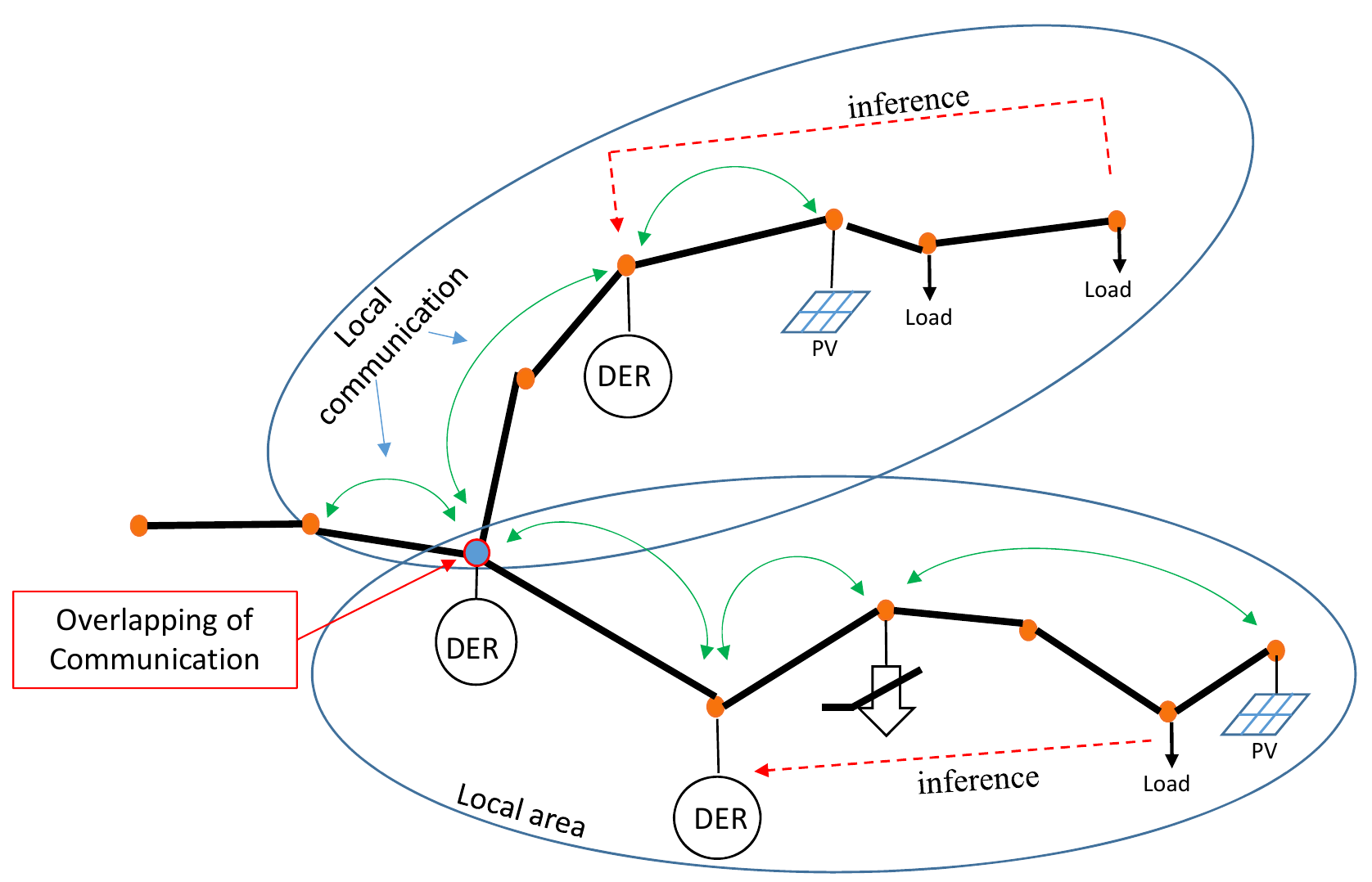}
        \caption{Two neighboring clusters}
        \label{fig:bus-cluster}
    \end{figure}

    In this section, a new distributed control regime is proposed to address the challenges identified in the previous sections. The proposed design assumes that local communications among neighbors are enabled and that, for the purpose of voltage inference, the distribution network is partitioned into overlapping clusters in terms of information collection and inferencing, as depicted in figure \ref{fig:bus-cluster}.
    
    The new regime consists of three distributed algorithms: (i) Distributed $Q$ controls that are implemented distributively; (ii) a distributed algorithm to determine the minimum and maximum voltages within the given distribution network; and (iii) distributed $P$ controls that coordinate active power injections in response to maximum/minimum voltage issues within the whole network.

\subsection{Distributed $Q$ Controls}
    
    The distributed $Q$ control aims to maintain the local voltages within the operational range by injecting reactive power. Specifically, the $Q$ control is to optimize the following performance index: if there is a DER at node $i$,
    \begin{equation}
        \min_{Q_{g_i}} \frac{1}{2}f_q^2(V_i-V_i^{ref}), \label{eq:fiv}
    \end{equation}
    where $V_i^{ref}$ is the voltage reference for node $i$, and function $f_q(\cdot)$ is the deadzone function defined by: for any chosen value $0\leq \sigma_v<<0.1$,
    \begin{equation}
        f_q(x) = \begin{cases}
        x,& \mbox{if } |x|\geq \sigma_v \\
        0,& \mbox{if } |x|< \sigma_v
        \end{cases}.
    \end{equation}
    Furthermore, voltage reference $V_i^{ref}$ is made adaptive by setting it to be the local average as:
    \begin{equation}
        V_i^{ref}=\text{SAT}_{[\underline{V}+\sigma_v,
        \overline{V}-\sigma_v]}\left(\lambda_i V_i +(1-\lambda_i)\frac{1}{|\clN_i|}\sum_{j\in \clN_i}V_j\right), \label{eq:Vref}
    \end{equation}
    where $|\clN_i|$ represents the cardinal of the neighboring set $\clN_i$, $\underline{V}$ and $\overline{V}$ are operational limits (e.g., 0.95 p.u. and 1.05 p.u., respectively), $\text{SAT}(\cdot)$ is the saturation function
    \begin{equation}
        \text{SAT}_{[\underline{v},\overline{v}]}(x) = \begin{cases}
        \underline{v},&x\leq\underline{v}\\
        x,&\underline{v}< x\leq \overline{v}\\
        \overline{v},&x>\overline{v}\end{cases},
    \end{equation}
    and $\lambda_i\in(0,1]$ is another design parameter.
    In \eqref{eq:Vref}, $V_j$ for $j\in\clN_i$ is either the real-time measurement sent by the communication neighboring  $j$th node or the inference value by the $i$th agent about the $j$th nodal voltage.

    Defining the reactive power utilization ratio as 
    \begin{align}
        \gamma_{q_i} \overset{\Delta}{=} \frac{Q_{g_i}}{\overline Q_{g_i}},\label{eq:alfaq}
    \end{align}
    we can choose distributed $Q$ control to be
    \begin{align}
            \dot \gamma_{q_i} =
            \begin{cases}
        0, & \gamma_{q_i}=1, \; \mbox{and} \; 
        h_{q_i} \geq 0; \\
        0 & \gamma_{q_i}=-1
        \; \mbox{and} \; h_{q_i} \leq 0; \\
        h_{q_i}(V_i,\gamma_{q_i},\gamma_{q_j}) & \mbox{else};
         \end{cases},
            \label{eq:dotxql}
    \end{align}
    where $k_{q_i}>0$ is the cooperative control gain, $0<\beta_i<<1$ is a stepsize, and 
    \begin{align} h_{q_i}(V_i,\gamma_{q_i},\gamma_{q_j})= & k_{q_i}\sum_{j \in { \clN}_i }(\gamma_{q_j}-\gamma_{q_i}) \nonumber \\
    & - \beta_i  (1-\lambda_i) \overline{Q}_{g_i}f_q(V_i-V_i^{ref}) \mathcal{X}_{ii}. \nonumber \end{align}
    In \eqref{eq:dotxql}, $\overline{Q}_{g_i}\equiv0$ where there is no DER, then $\gamma_{q_i}$ will only collect the average utilization ratio information and pass it to the neighbors.
    
    It follows from \eqref{eq:dotxql} that, if $V_i>V_i^{ref}$ and $\gamma_{q_i}$ is not close to negative one, $\gamma_{q_i}$ will become more negative, that is, more reactive power will be injected to push down the local voltage. Analogously, more reactive power will be absorbed if  $V_i<V_i^{ref}$ and $\gamma_{q_i}$ is not one yet. 
    Should design parameter $\sigma_v$ be chosen to be a small positive value, deadzone function $f_q(\cdot)$ prevents excessive reactive power injection that could be needed to match the local voltage exactly to its reference. 
    
    It will be shown in sections \ref{sec:result} and \ref{sec:simu} that $Q$ control \eqref{eq:dotxql} ensures local consensus of $\gamma_{q_i}$. Nonetheless, the desired voltage profile $V_i^{ref}$ varies between $0.95$ to $1.05$ across the distribution network, $\gamma_{q_i}$ will vary between $-1$ and $1$ across the network, which is different from \cite{xin2011self,xin2011cooperative,maknouninejad2014realizing}.
    Another main difference is that the limits of reactive power control are explicitly taken into account in \eqref{eq:dotxql}.

\subsection{Distributed Estimation of Maximum and Minimum Voltages}

Except for battery banks, a majority of DERs including  PVs can scale down active power injection, often referred to as curtailment, but the power that was reduced becomes a pure loss. Hence, the goal of active renewable generation is to inject as much active power as generated, provided that voltage stability is maintained throughout the distribution network. 

To enable intelligence and autonomous $P$ controls that minimizes curtailments in distribution networks with high penetration renewable generation, it is critical for every distributed controller to gain such critical real-time knowledge as the maximum and minimum voltages (defined as $V^h$ and $V^l$) in the system. This is done by using the following cooperative protocols: if there is a DER at the $i$th node,
\begin{eqnarray}
&& \hspace*{-0.45in} V_i^{h}(t) = \max\left\{ \max_{j\in \clN_i} V_j(t), \; (1-\Delta t)\max_{j\in \clN_i}  V_j^h(t-\Delta t)\right\}, \label{eq:max} \\
&& \hspace*{-0.45in} V_i^{l}(t)  =  \min\left\{ \min_{j\in \clN_i} V_j(t), \; (1-\Delta t)\min_{j\in \clN_i} V_j^l(t-\Delta t)\right\}, \label{eq:min}
\end{eqnarray}
where $V_i^{h}$ and $V_i^{l}$ are the node-$i$'s estimates of maximum and minimum voltages in the network, respectively, $\Delta t>0$ is a small time increment.  In \eqref{eq:max} and \eqref{eq:min}, for node $j\in\clN_i$ that does not have a DER, $V_j$ is the inference value by the $i$th agent, and $V_j^l=V_j^h=V_j$.

\begin{figure}[!h]
            \centering
                \includegraphics[width=0.75 \linewidth]{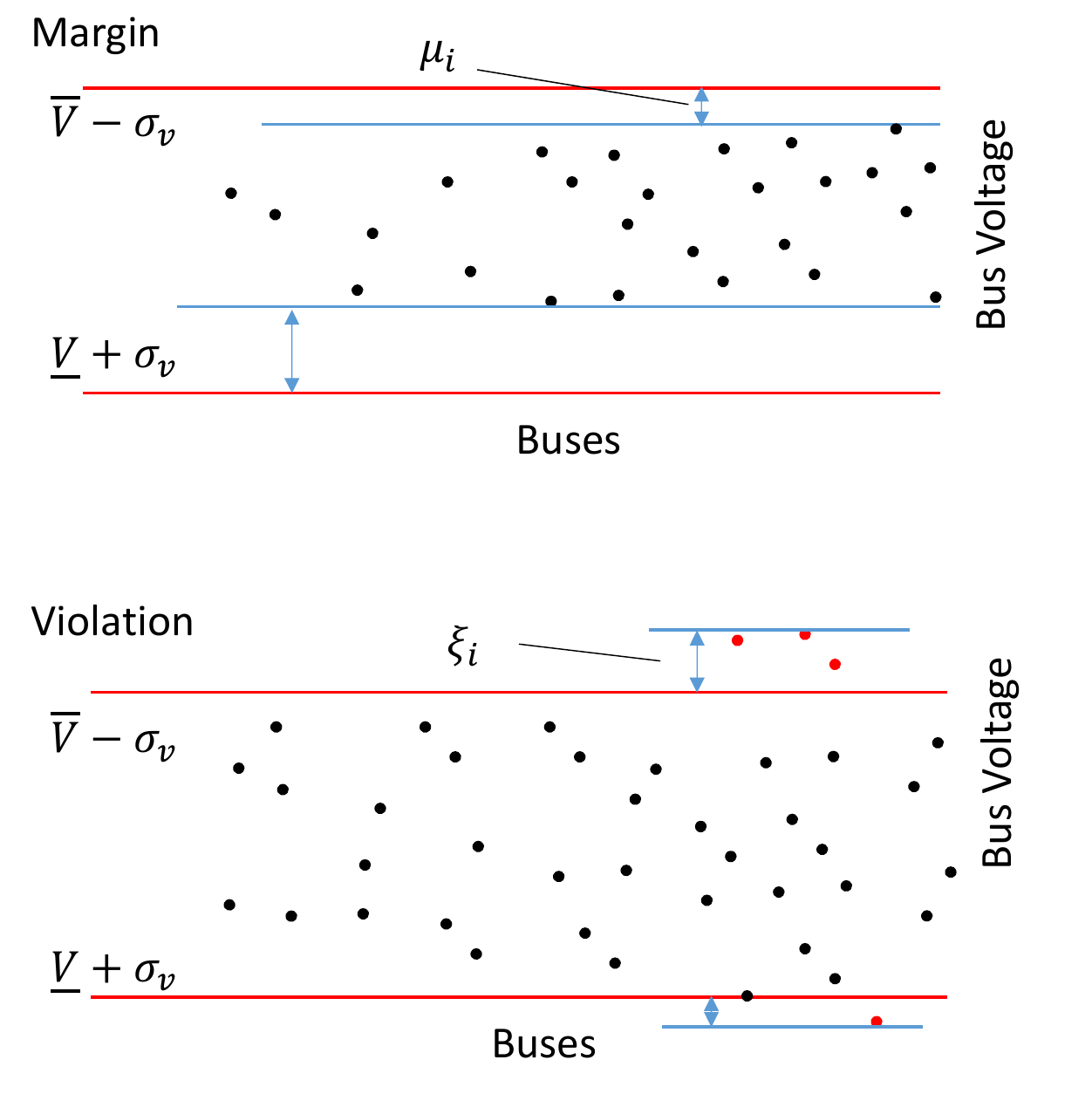}
                \caption{Voltage stability margin and  voltage violation index.}
                \label{fig:muk}
    \end{figure}

Based on the estimates $V_i^{h}$ and $V_i^{l}$, each of DERs can estimate the network's voltage stability margin
$\mu$ as:
    \begin{align}
    \mu_i = & \min\left\{  \max\{ V_i^l-(\underline{V}+\sigma_v), \; 0\},\right.\nonumber
                \\& \hspace*{.4in}
                \max\{ (\overline{V}-\sigma_v)-V_i^h, \; 0\}\Big\}, \label{eq:margin}
    \end{align}
or equivalently the network's voltage violation index $\xi$ as:
        \begin{align}
                \xi_i = & \max\left\{ \max\{(\underline{V}+\sigma_v)- V_i^l, \; 0\},\right.\nonumber
                \\& \hspace*{.4in} 
                \max\{ V_i^h -(\overline{V} -\sigma_v), \; 0\} \Big\}. \label{eq:violation} 
            \end{align}
 It is clear that $\mu_i>0$ implies $\xi_i=0$ and that $\xi_i>0$ yields $\mu_i=0$. It is possible that $\xi_i=\mu_i=0$, and their zero values are the boundary between having voltage stability margin and suffering voltage violation. These two complementary measures are illustrated graphically by figure \ref{fig:muk}. 

\subsection{Distributed and $\overline{Q}$-Coordinated $P$ Controls}

For non-storage DERs, their $P$ controls aim to maximize their active power injection (or minimize the curtailment) as long as the voltage stability is maintained. To this end, lets define the {\it curtailment ratios}: for all node $i$ where there is a DER, 
    \begin{align}
        C_{p_i} \overset{\Delta}{=} 1- \frac{P_{g_i}}{\overline P_{g_i}}.\label{eq:curtailment}
    \end{align}
Hence, the $P$-control objective can be expressed as:
    \begin{align}
         & \textbf{whenever}\;\; \xi_i>0, \;\; 1-|\gamma_{q_i}|<\epsilon_q,\nonumber \\        
        & \min_{P_{g_k}} C_{p_i},  \label{cstrtv}\\
        & \textbf{such that}\;\; \mu_i>0,\nonumber
    \end{align}
    where $\epsilon_q>0$ is a small threshold value. If inequality $1-|\gamma_{q_i}|<\epsilon_q$ holds, it means that reactive power control capacity is exhausted. As explained in the previous subsection, $\xi_i>0$ implies that there is a voltage violation somewhere in the network. When both conditions hold, reactive power capacity in the network is insufficient to handle the voltage problem, and curtailment has to be introduced to address the issue, and the formulation \eqref{cstrtv} is to find the minimal curtailment toward meeting voltage operational requirements.
   
    To solve the conditional optimization problem \eqref{cstrtv}, we propose the following distributed cooperative $P$ control:
    \begin{align}
        \dot C_{p_i} =
            \begin{cases}
        0, & C_{p_i}=1, \; \mbox{and} \; 
        h_{p_i} \geq 0; \\
        0 & C_{p_i}=-1
        \; \mbox{and} \; h_{p_i} \leq 0; \\
        h_{p_i}(\xi_i,\mu_i,C_{p_i},C_{p_j}) & \mbox{else};
         \end{cases},
            \label{eq:dotxp}
    \end{align}
    where $\beta_i^{\prime}> 0$ is a stepsize, and $k_{p_i}>0$ is the cooperative control gain, and 
    \begin{align} 
        & h_{p_i}(\xi_i,\mu_i,C_{p_i},C_{p_j}) \nonumber \\
        =& k_{p_i} \sum_{j\in\clN_i}  (C_{p_j}-C_{p_i}) +\beta_i^{\prime} \max\{ 1-|\gamma_{q_i}|,\; \epsilon_q\} \xi_i \nonumber\\
            &- \beta_i^{\prime} \min\{ 1-|\gamma_{q_i}|, \; \epsilon_q\}\mu_i. \label{eq:hp}
    \end{align}
    The control in \eqref{eq:dotxp} and \eqref{eq:hp} contains a consensus protocol as its first term so that curtailment if required is fairly distributed within the network to ensure voltage stability. The second term, $ \beta_i^{\prime} \max\{ 1-|\gamma_{q_i}|,\; \epsilon_q\} \xi_i$ ensures that, if there is voltage violation but the reactive power capacity is used up, curtailment either is introduced or increased to relieve the voltage problem. And, the third term, $ \beta_i^{\prime} \min\{ 1-|\gamma_{q_i}|, \; \epsilon_q\}\mu_i$ is to reduce the curtailment when there is no longer any voltage issue. Both the second and third terms are subgradient based control components, and they do not fight each other due to the complimentary nature of measure $\mu$ and index $\xi$ as discussed before.
    
\subsection{Main Results} \label{sec:result}

The main results in this paper are summarized into the following theorem whose proof is included in appendix \ref{sec:proof}.
    \begin{theorem}
    \label{thm_sp}
    The proposed control regime has following properties: \\
    (i) Distributed $Q$ control \eqref{eq:dotxql} optimizes objective function \eqref{eq:fiv} within the constraints of $|Q_{g_i}|\leq \overline{Q}_{g_i}$ till $f_q(V_i-V_i^{ref})=0$. And, it ensures local consensus of $ \gamma_{q_i}$. \\
    (ii) Distributed protocols \eqref{eq:max} and \eqref{eq:min} provide real-time information about maximum and minimum voltages in the network. \\
    (iii) Under bidirectional, local and connected (possibly varying) communication among DERs, distributed $P$ control \eqref{eq:dotxp} minimizes curtailments while working with cooperative $Q$ control to ensure voltage stability.
    \end{theorem}

\section{Simulation} \label{sec:simu}

In this paper, a three-layer
software platform is developed: the physical layer containing a distribution system simulating engine
(based on OpenDSS \cite{OpenSourcePlat}) with an explicit input-output DER model, the communication layer, and the algorithm and control layer\footnote{The open source software entitled as MA-OpenDSS and all the test circuits are available on \url{https://www.ece.ucf.edu/~qu/ma-opendss}}.
These three-layer design enables a set of distributed optimization and
control algorithms including the proposed control regime. The performance of the proposed method has been evaluated on multiple test circuits such as: IEEE 123, 8500-node, and other larger systems. 

\subsection{IEEE 123 system}
\begin{figure}
    \centering
    \includegraphics[width=0.85\linewidth]{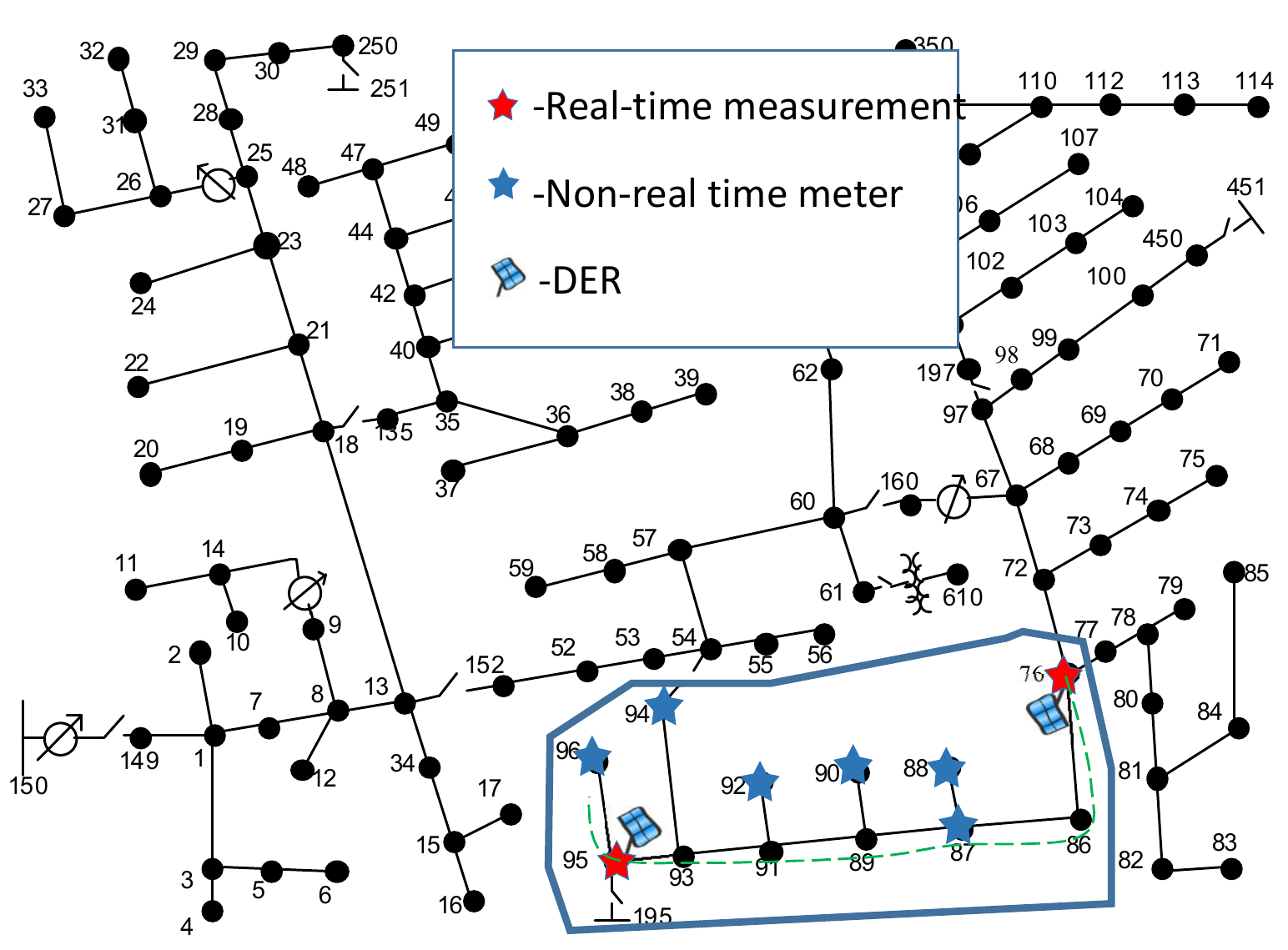}
    \caption{IEEE 123 system with DERs in a cluster.}
    \label{fig:123}
\end{figure}

\begin{figure}[ht]
    \centering
    \includegraphics[width=0.8\linewidth]{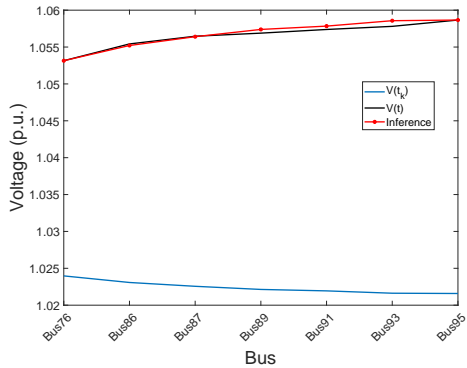}
    \caption{Voltage inference at the DER bus in the IEEE 123 system.}
    \label{fig:infer}
\end{figure}

First, the voltage inference method is illustrated on IEEE 123 system. In the test case, assume a communication cluster has been established as encircled in figure \ref{fig:123}. The PV buses have real-time measurement and the rest buses in the cluster have regular digital meters which can offer 15-minute measurements. Assume the load is decreasing right after the peak hours and the DERs are changing ($t_k\to t$). The predicted voltages of non-real time measured buses at time $t$ are shown in figure \ref{fig:infer}, which match the the real values $V(t)$ well (about $1\%$ error) with a big change of the operating condition, i.e. a injection increase of $40\%$ of total load at bus $\#95$, a decrease of $20\%$ at bus $\#72$, and $20\%$ drop of the load.

\subsection{IEEE 8500 system}
\begin{figure}[ht]
    \centering
    \includegraphics[width=\linewidth]{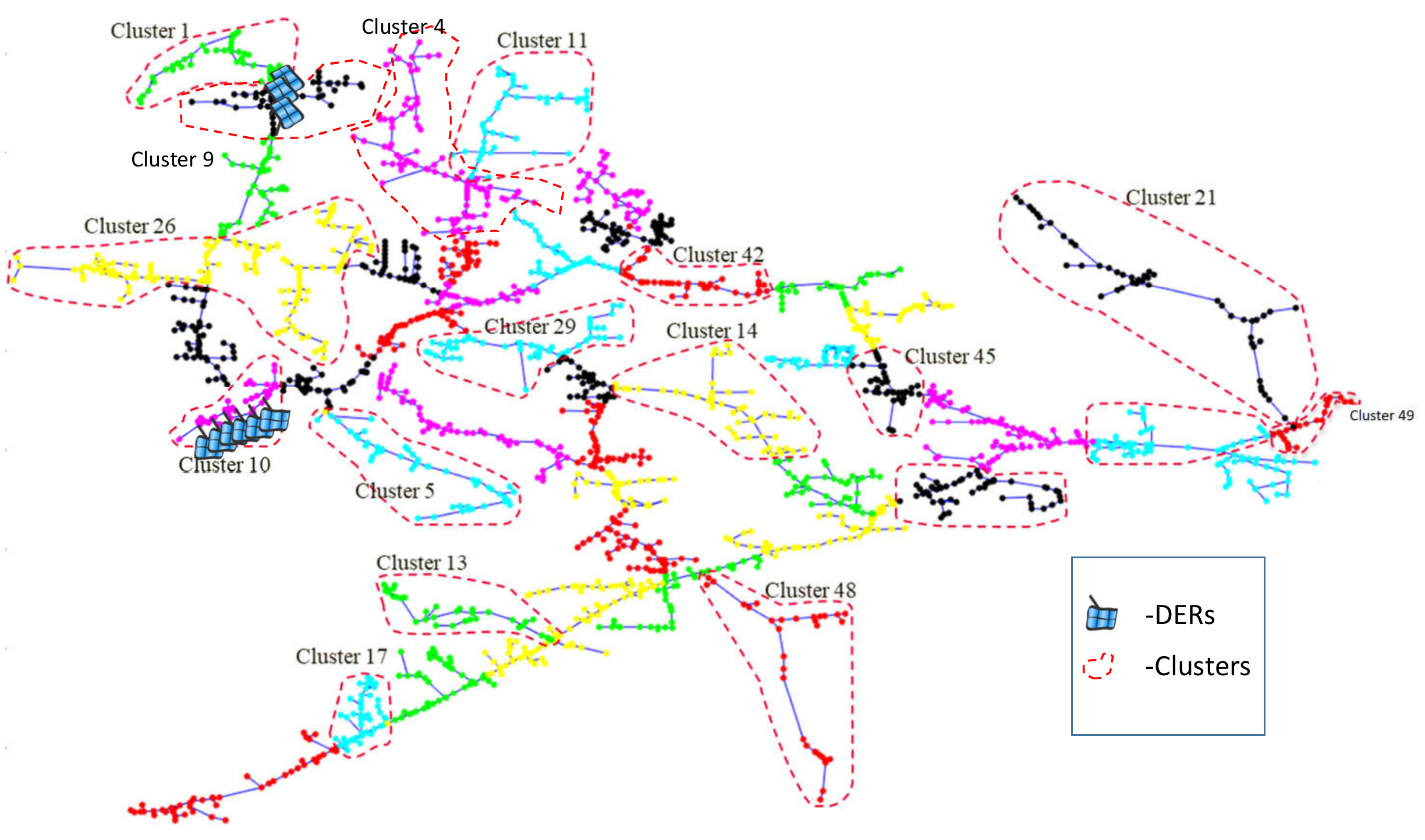}
    \caption{DERs and clusters defined in a case test of the IEEE 8500-node system.}
    \label{fig:clstrs}
\end{figure}

The IEEE 8500-node system is used to test the proposed cooperative $Q$ controls and $\overline Q$-Coordinated P controls. The communication structure is predefined for IEEE 8500 system using the software platform, which has 49 communication clusters, as shown \ref{fig:clstrs}. A $100\%$ penetration case \cite{ecc2020} with 10 large-scale PVs installed in clusters 9 and 10 is studied. Given at least one node in each cluster has real-time measurement, so the voltage inference algorithm is implemented for the overall system. 

\begin{figure}[ht]
    \centering
    \includegraphics[width=0.75\linewidth]{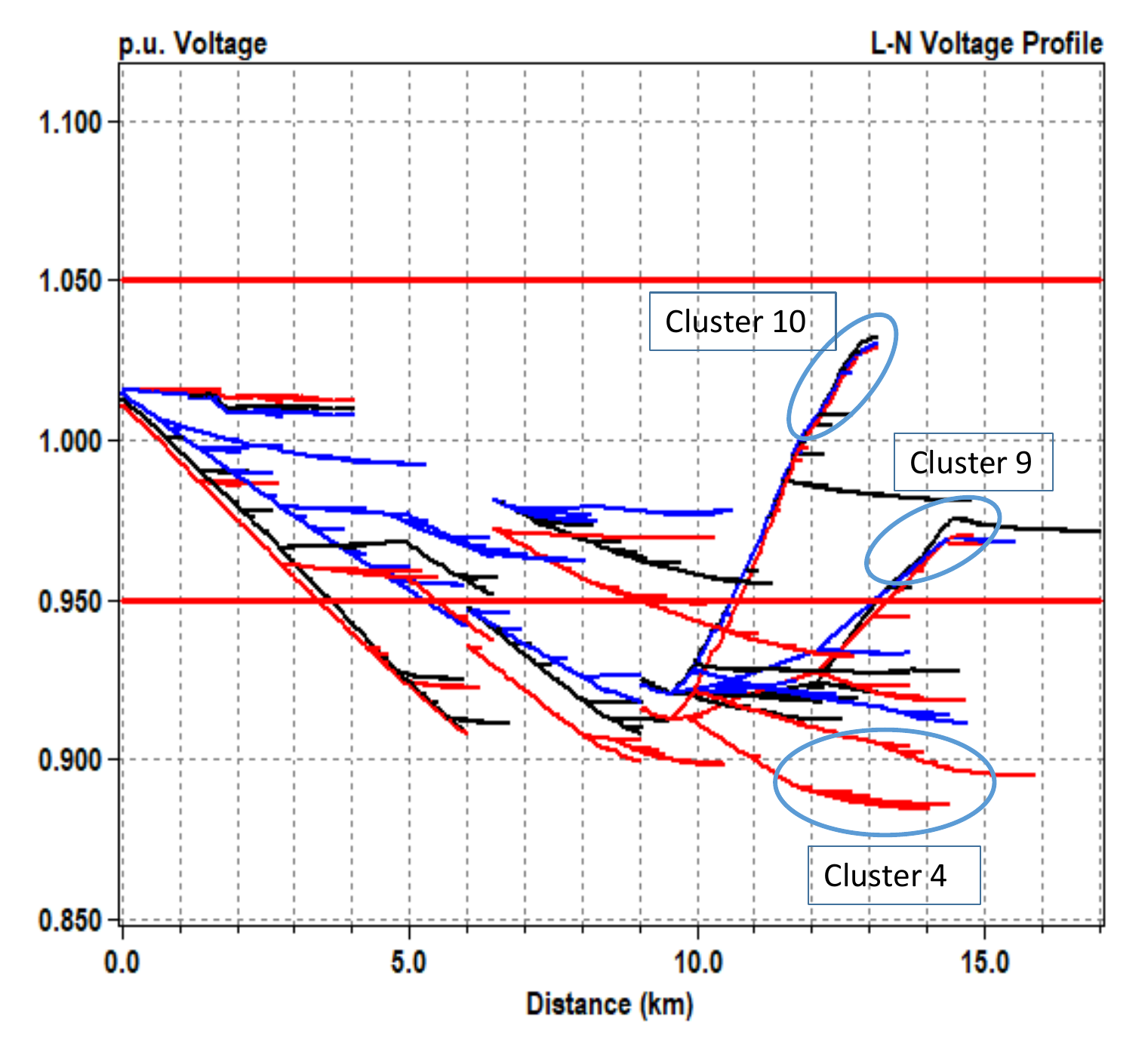}
    \caption{Voltage violation when reactive power control is performed only.}
    \label{fig:lowvolt}
\end{figure}

First, distributed Q control is tested. Since the spatial power unbalances, in this case, is obese between power sources and loads, the voltage deviations are enormous, which can not be mitigated by merely reactive power control, as shown in figure \ref{fig:lowvolt}. 

\begin{figure}[ht]
    \centering
    \includegraphics[width=0.75\linewidth]{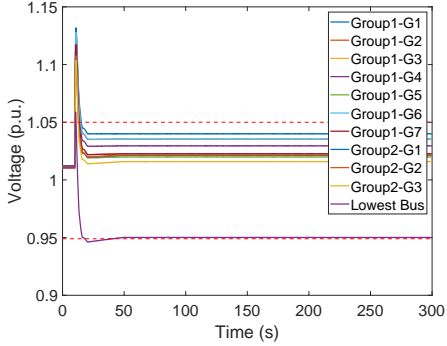}
    \caption{Voltages monitored in IEEE 8500-node system.}
    \label{fig:85v}
\end{figure}

\begin{figure}[ht]
    \centering
    \includegraphics[width=0.75\linewidth]{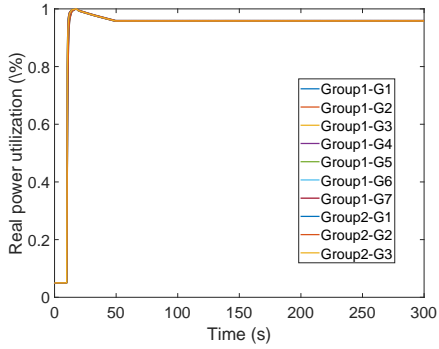}
    \caption{Real power utilization ratios of DERs.}
    \label{fig:85alfap}
\end{figure}

\begin{figure}[ht]
    \centering
    \includegraphics[width=0.75\linewidth]{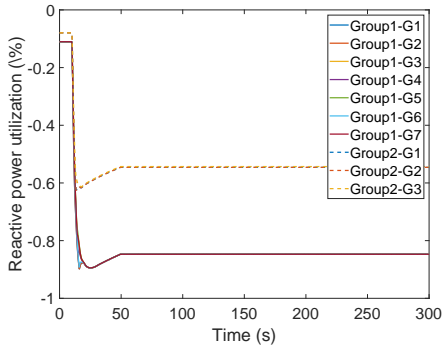}
    \caption{Reactive power utilization ratios of DERs.}
    \label{fig:85alfaq}
\end{figure}

The control response of the proposed $\overline Q$-coordinated P control is shown in figures \ref{fig:85v}, \ref{fig:85alfap}, and \ref{fig:85alfaq}. 
As the results show, the system voltages increase due to the power injection at $t=10s$, then the local reactive power control starts to push the voltage down. At around $t=12s$, the voltage at the lowest bus starts to violate the lower limit of voltage regulation. This is because the local reactive power control is only for the local voltage, shown in figure \ref{fig:85v}. At the same time, the P control starts to kick in, and the real power control (curtailment) responds to the system-level objective. Eventually, the control settles down, and all the voltages throughout the network are well controlled. Moreover, the consensus of real power utilization ratios ($\gamma_{p_i}$) is achieved throughout the system, shown in figure \ref{fig:85alfap}; and reactive power utilization ratios ($\gamma_{q_i}$) converge to one value for each cluster respectively, shown in figure \ref{fig:85alfaq}. Note that the software developed is capable of dealing with communication delays, the delays have been considered in this test: 0.1s delay on node-to-node and 0.8s delay on cluster-to-cluster communications.

\subsection{Synthetic 100k-system}
\begin{figure}[!ht]
\centering
         \includegraphics[width=0.75\linewidth]{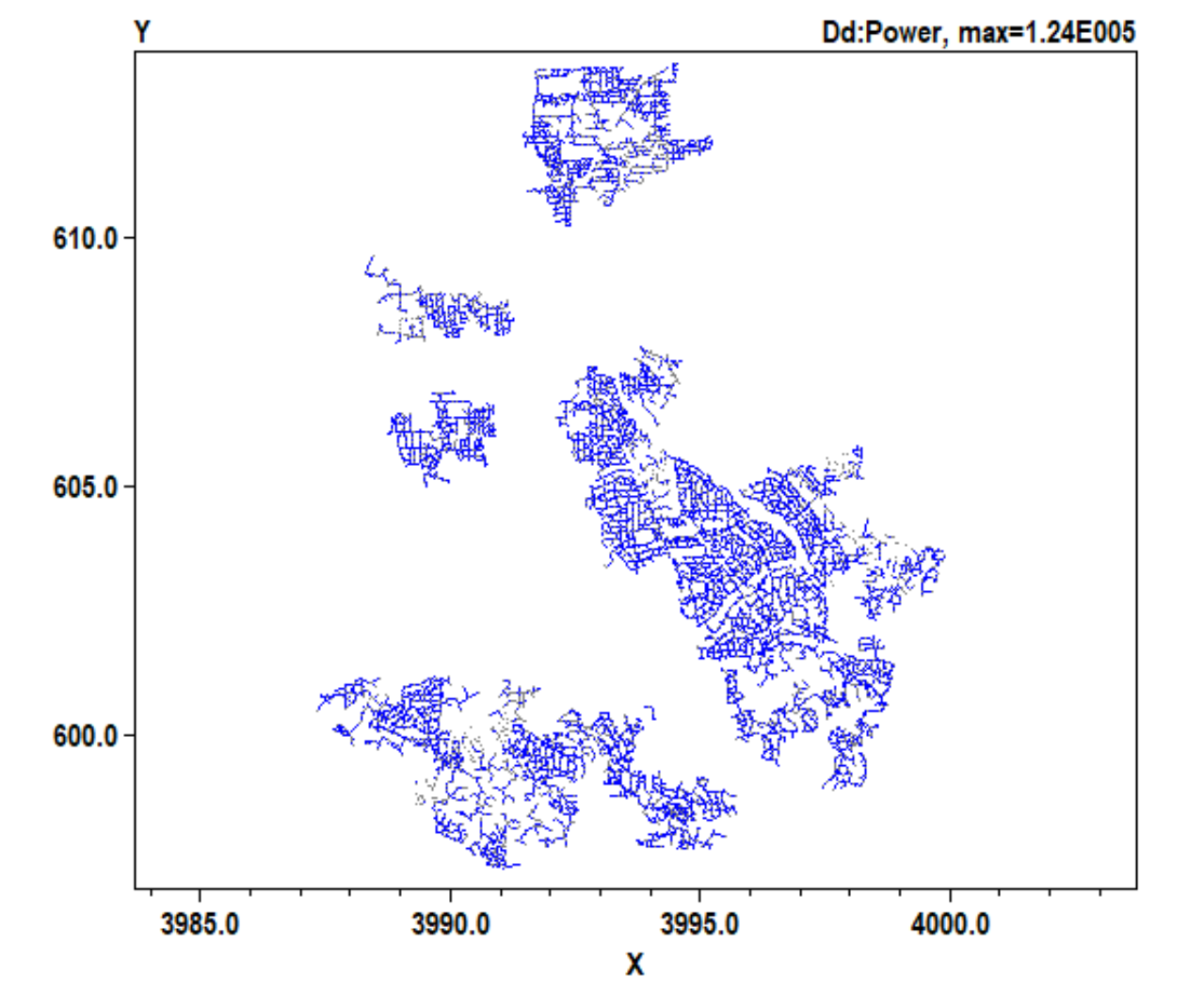}
    \caption{The 100-k synthetic circuit.}
    \label{fig:C1ckts}
\end{figure}

The effectiveness of the proposed approach on larger-scale systems is also tested on a 100-k circuit built by combining different types of feeders: the circuit is assembled by 12 urban/suburban feeders, as shown in figure \ref{fig:C1ckts}. The circuit has 75392 devices, 53011 buses, 108848 nodes, and 124MW+j47.3MVar loads in total. 
In the communication layer, 268 clusters are defined for communication and control.

\begin{figure*}[!ht]
\centering
\subfloat[Voltage profile without control]{
         \includegraphics[width=0.29\linewidth]{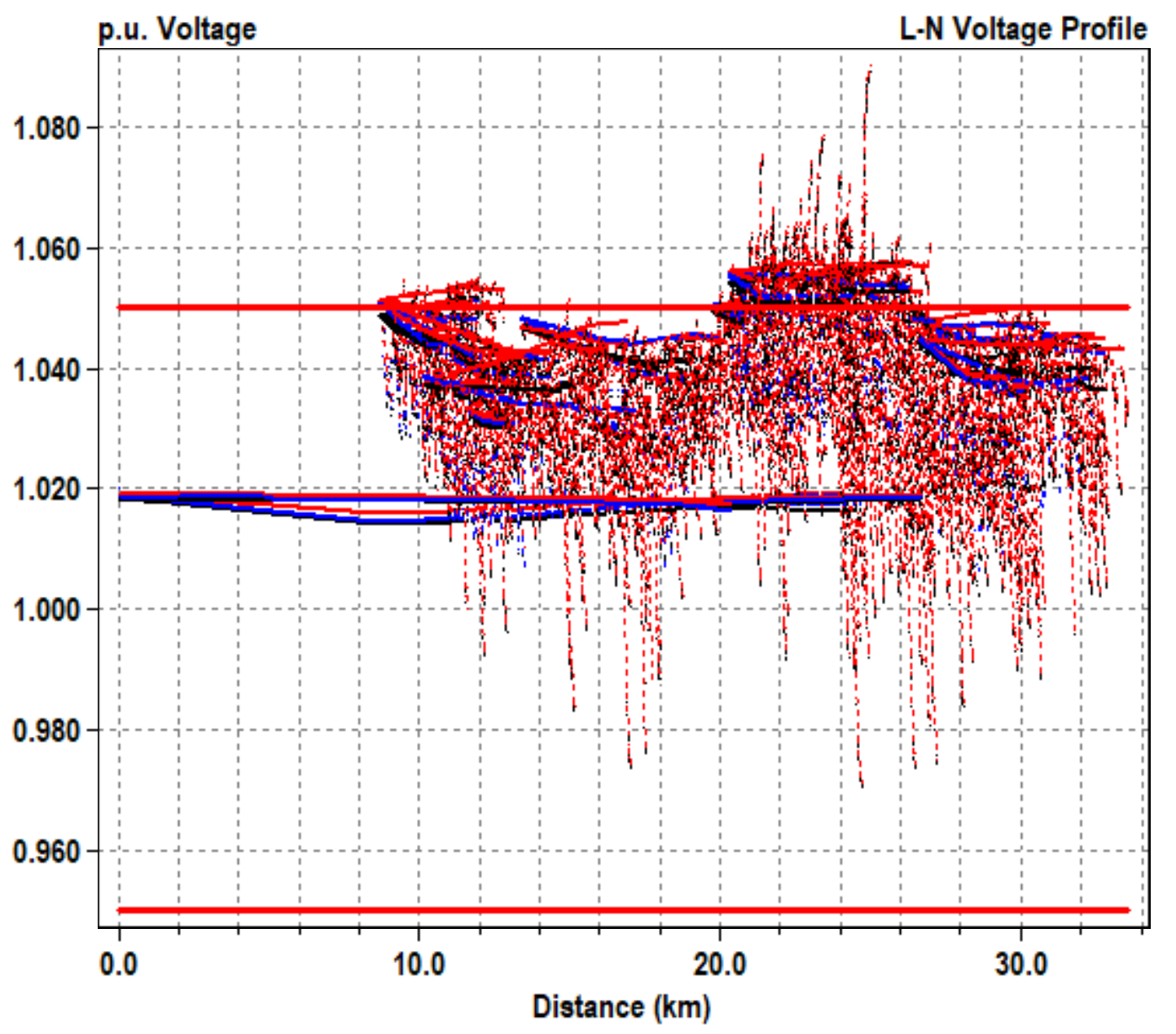}
         \label{fig:100knocontrol}
         }
\subfloat[Result of the distributed Q control]{
         \includegraphics[width=0.29\linewidth]{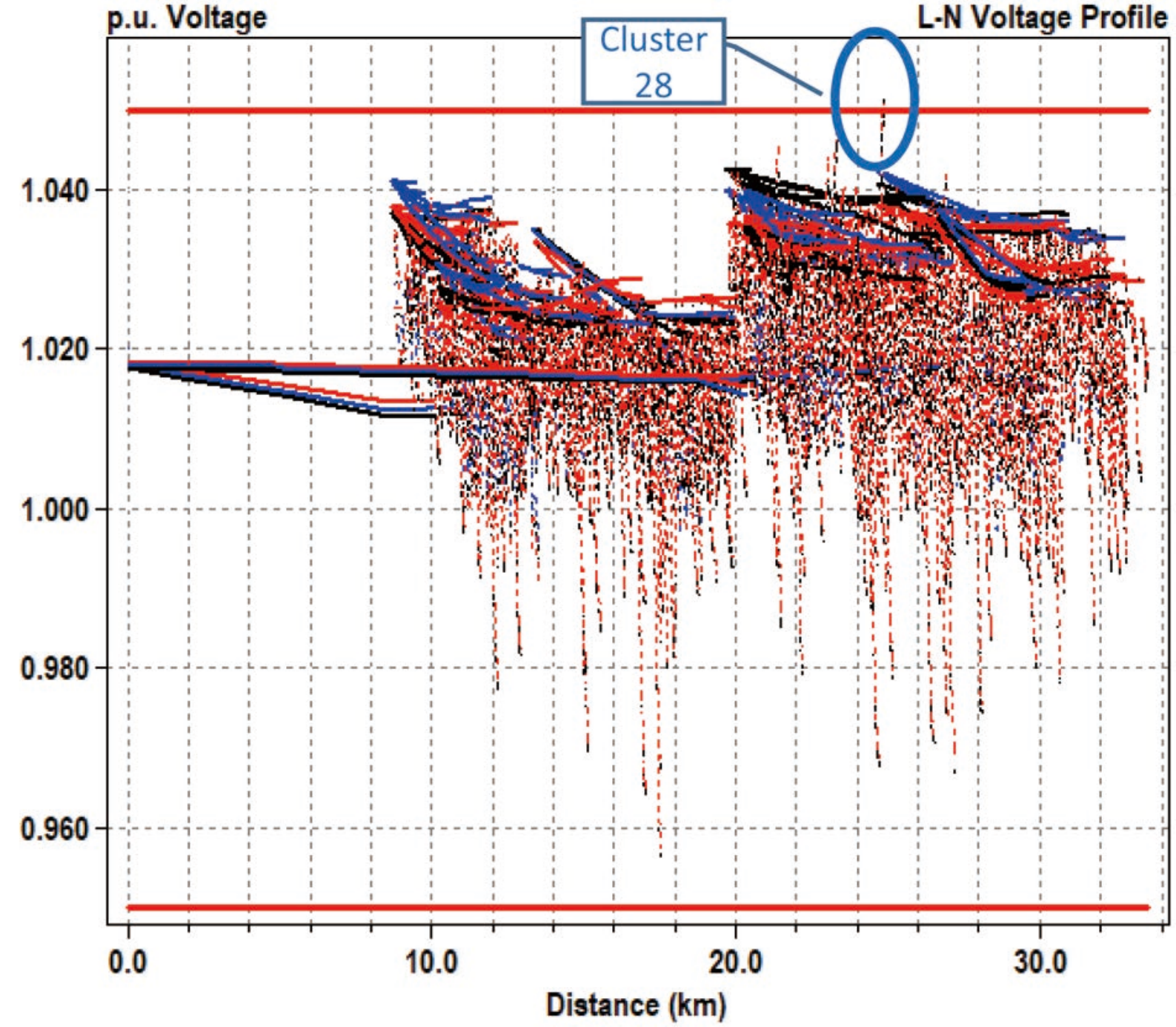}
         \label{fig:C1cktscase2}
         }
\subfloat[Detailed information in cluter 28]{
         \includegraphics[height = 1.9in,width=0.41\linewidth]{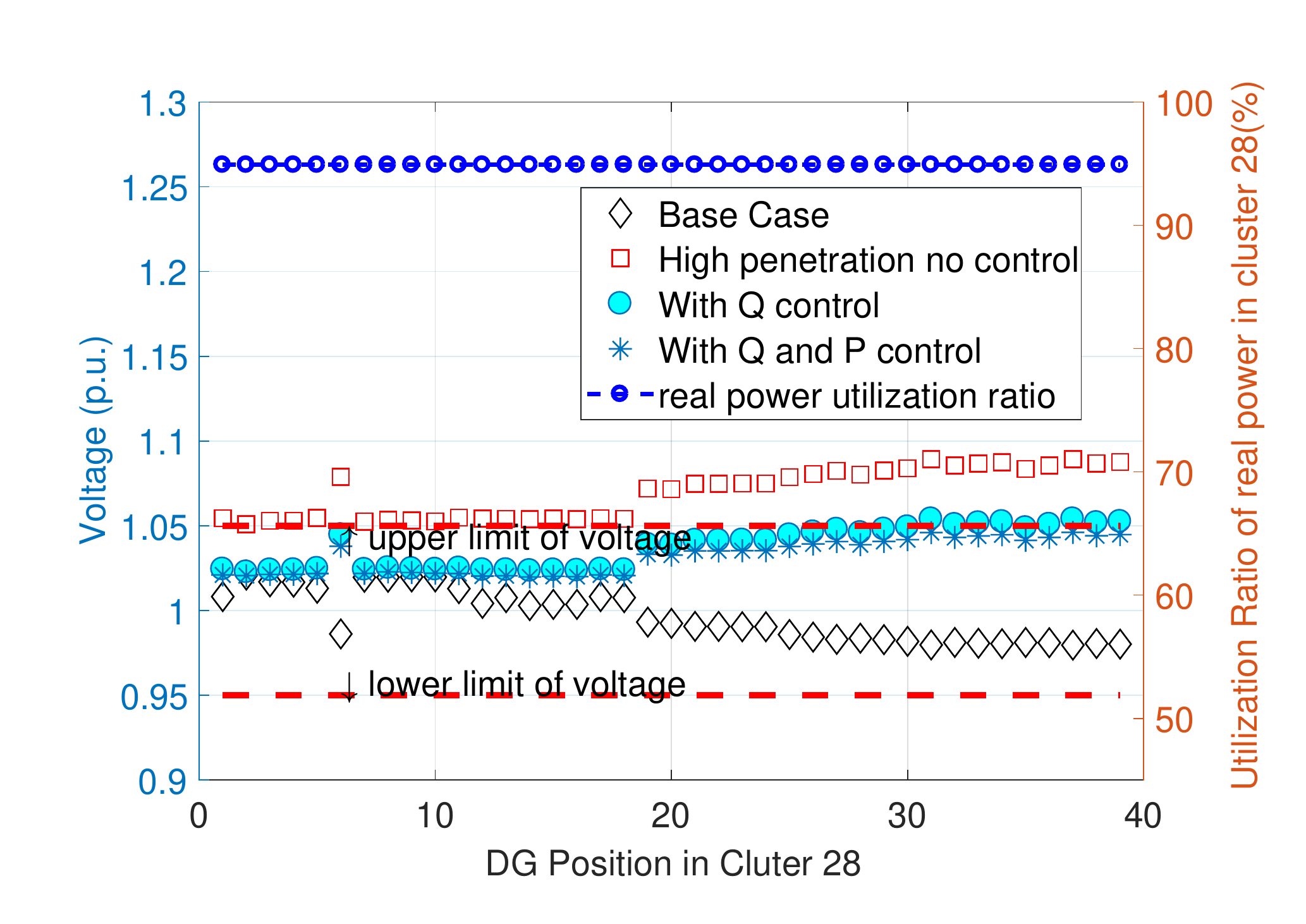}
         \label{fig:bsvcase}
         }
    \caption{The control results in Scenario 2 of 100-k system.}
    \label{fig:C1curt}
\end{figure*}

Using a greedy search method \cite{RATHBUN2018}, a $100\%$-penetration scenario is founded to have 42,146 PVs distributively deployed in the system. Figure \ref{fig:100knocontrol} shows the voltage violation of this case. 
Simulation results show that the proposed cooperative reactive power control can effectively regulate the system voltages except the buses in cluster 28, as shown in Figure \ref{fig:C1cktscase2}.

Then the real power curtailment is enforced on PVs to mitigate the violations. As shown in figure \ref{fig:bsvcase}, the $\overline{Q}$-coordinated P control introduces a 4\% curtailment as supplementary voltage control, so that all bus voltages are well controlled, shown by the stars, and the power utilization ratios converge as well, shown by the blue circle-dot line.

\section{Conclusion}
The fully distributed voltage control strategy is proposed for large-scale power distribution systems with high-penetration DERs, which has two levels: the distributed Q control and the $\overline{Q}$-coordinated P control. The voltage inference method is derived from the DER-explicit distribution network model and its voltage sensitivities, which is capable of enhancing observability of grid-edge areas.
All the algorithms have been tested in either standard or synthetic large systems. Moreover, an open-source software platform is developed, which has general modelling of the information, control, and physical systems, and can be used to test the distributed control and optimization algorithms in large-scale distribution systems.

\appendices
\renewcommand{\theequation}{\thesection.\arabic{equation}}
\setcounter{equation}{0}

\section{A Model of distribution networks}\label{appen:model}

For any radial distribution network shown in figure \ref{fig:busI}, the following branch model \cite{Low2013branch} is standard:
\begin{subequations}
\begin{align}
 P_{s_i}  - P_i =& \; \sum_{m \in \mathcal{C}_i} (P_{s_m}+ R_m \ell_m),  \label{eq:brcha} \\
 Q_{s_i}  - Q_i = & \; \sum_{m \in \mathcal{C}_i} (Q_{s_m}+X_m \ell_m), \label{eq:brchb} \\
V_{\Gamma_i}^2 - V_i^2 = & \; 2(R_i P_{s_i} +X_i Q_{s_i})+{|z_i|}^2\ell_i, \label{eq:brchc} \\
{V_{i}}^2 \ell_i = & \; {P_{s_i}}^2 + Q_{s_k}^2, \label{eq:brchd}
\end{align}
\label{eq:brchflow}
\end{subequations}
where $\ell_i = |\tilde I_i|^2$, and $z_i=R_i+jX_i$ is its impedance of the line from $\Gamma_i$ to bus $i$. It follows from \eqref{eq:brchflow} that
\begin{align}
V_{\Gamma_i}^2-V_{i}^2  &=2P_{s_i}R_i+2 Q_{s_i}X_i+P_{L_i}R_i+Q_{L_i}X_i,\label{eq:vii-1}
\end{align}
\noindent where $(P_{L_i}+jQ_{L_i})$ is the complex loss over the line from from $\Gamma_i$ to bus $i$ as
\begin{equation}
P_{L_i} = {(P_{s_i}^2+Q_{s_i}^2)R_i}/{V_{i}^2}, \;\;\;
    Q_{L_i}={(P_{s_i}^2+Q_{s_i}^2)X_i}/{V_{i}^2}. \label{eq:plpsqs}
\end{equation}
Equation \eqref{eq:plpsqs} shows that $Q_{L_j}=b_jP_{L_j}$ with $b_j=X_j/R_j$. At any end bus (say, $n$), $P_{s_n} = P_n$ and $Q_{s_n}=Q_n$.

Given bus set $\mathcal{C}_i$, the energy conservation equation at bus $i$ can be written as
\begin{equation}
 P_{s_i} = P_i+\sum_{m\in \mathcal{C}_i}(P_{m}+P_{L_m}), \; Q_{s_i} = Q_i + \sum_{m\in \mathcal{C}_i}(Q_{m}+b_m P_{L_m}).\label{eq:psqs}
\end{equation}
Substituting \eqref{eq:psqs} into \eqref{eq:vii-1} yields
\begin{align}
V_{\Gamma_i}^2-V^2_{i} = & \; 2\sum_{m\in \mathcal{C}_i\cup\{i\}}( R_i P_m + X_i Q_{m})\nonumber\\
& \; +2\sum_{m\in \mathcal{C}_i}(R_i+b_mX_i)P_{L_m} \nonumber\\
& \; +(R_i+b_iX_i)P_{L_i}. \label{eq:exact}
\end{align}
Letting $i=1,\cdots,k$ in \eqref{eq:exact} and adding all these equations together yield the concise network equation \eqref{eq:vpl} in section II.A. Substituting \eqref{eq:psqs} and \eqref{eq:businj} into \eqref{eq:brchd} and using the definition of $P_{L_k}$ yield \eqref{eq:plv}.

\section{Proof of Theorem \label{sec:proof}}

    \begin{proof}
    (i) It is obvious that \eqref{eq:fiv} is convex and that its subgradient with respect to the control design variable $\gamma_{q_i}$ is
    \begin{eqnarray*}
    \frac{\partial  f^2_q(V_i-V_i^{ref})}{\partial \gamma_{q_i}}
    & = &  \frac{\partial  f^2_q(V_i-V_i^{ref})}{\partial V_i}
    \frac{\partial V_i}{\partial \gamma_{q_i}} \\
    & = & (1-\lambda_i)f_q(V_i-V_i^{ref}) \frac{\partial V_i}{\partial \gamma_{q_i}}. 
    \end{eqnarray*}
    It follows from \eqref{eq:fiv} and \eqref{eqvklnr}, and \eqref{eq:alfaq} that
    \begin{align}
    \frac{\partial  f^2_q(V_i-V_i^{ref})}{\partial \gamma_{q_i}}
    =  (1-\lambda_i)f_q(V_i-V_i^{ref})\overline{Q}_{g_i}
    \mathcal{X}_{ii}. \nonumber 
    \end{align}

    (ii) The proof is analogous to that of lemma 6.22 on page 280 in \cite{qu2009cooperative}. The weight $(1-\Delta t)$ is a forgetting factor so the maximum and minimum are the current values rather than historic values. 
    
       (iii) 
        It follows from (ii) that $V^h_i$, $V^l_i$, $\xi_i$ and $\mu_i$ acquire their global values of $V^h$, $V^l$, $\xi$, and $\mu$, respectively. For simplicity, lets consider in the rest of the proof \eqref{eq:dotxp} and \eqref{eq:hp} while  $\xi_i$ and $\mu_i$ are assumed to have converged to $\xi$ and $\mu$, respectively. The more general proof can be done analogously by considering not only \eqref{eq:dotxp} and \eqref{eq:hp} but also \eqref{eq:max}, \eqref{eq:min}, \eqref{eq:margin} and \eqref{eq:violation} all together.
        
        Lets denote the instantaneous maximum and minimum curtailment values in the distribution network:
        \[ C_{max}(t) = \max_i C_{p_i}(t), \;\;\; C_{min}(t) = \min_i C_{p_i}(t),   \]
        where $C_{p_i}(t)$ is defined by \eqref{eq:curtailment}. It follows from \eqref{eq:dotxp} and \eqref{eq:hp} that $0\leq C_{min}(t) \leq C_{max}(t) \leq 1$. 
        
        When there is no DER injection (which also corresponds to $C_{max}(t)=0$), the distribution network does not have any voltage problem (that is, $\mu>0$ and $\xi=0$). As the penetration level increases, the voltage problem may arise and, if so, reactive power compensation kicks in. Should reactive power control be insufficient to keep voltages within the allowable range, curtailment will be activated across the network (i.e., $C_{min}(t)$ would become positive). Nonetheless, there is a minimum curtailment level $C^*\in(0,1]$ at which $\mu=0$ and $\xi=0$ if $C_{p_i}=C^*$ for all $i$. This means that, when $C_{max}(t)<C^*$, there is either over-voltage or under-voltage or both ($\mu=0$ and $\xi>0$) and that,
        when $C_{min}(t)>C^*$, there is a voltage margin ($\mu>0$ and $\xi=0$).
        
        Assume without loss of any generality that, at time $t$, there are sets ${\cal N}_{max}(t)$ and ${\cal N}_{min}(t)$ such that $C_{p_k}(t) = C_{max}(t)$ for $k\in {\cal N}_{max}(t)$ and 
        $C_{p_l}(t) = C_{min}(t)$ for $l\in {\cal N}_{min}(t)$. It follows from \eqref{eq:dotxp} and \eqref{eq:hp} that 
        \begin{eqnarray} 
        \frac{1}{k_{p_i}}\dot C_{p_i} & = &  \underbrace{\sum_{j\in\clN_i}  (C_{p_j}-C_{p_i})}_{f_{c_i}} +\underbrace{\frac{\beta_i^{\prime}}{k_{p_i}} \max\{ 1-|\gamma_{q_i}|,\; \epsilon_q\} \xi_i}_{u_{i1}} \nonumber\\
        & & - \underbrace{\frac{\beta_i^{\prime}}{k_{p_i}} \min\{ 1-|\gamma_{q_i}|, \; \epsilon_q\}\mu_i}_{u_{i2}}. 
        \label{eq:agent-i}
        \end{eqnarray}
        Dynamics of equation \eqref{eq:agent-i} are linear with respect to "exogenous" inputs $u_1$ and $u_2$. Defining the network's weighted average curtailment 
        \[ M(t) = \frac{1}{N_g}\sum_{i=1}^{N_g} \frac{1}{k_{p_i}}C_{p_i}, \]
        we know from DERs having bidirectional communication that 
        \begin{equation} \dot{M}= \frac{1}{N_g} \sum_{i=1}^{N_g} (u_{i1} - u_{i2}), \label{eq:average}
        \end{equation}
        where $N_g$ is the total number of DERs. Therefore, the following three conclusions can be concluded from
        \eqref{eq:agent-i} and \eqref{eq:average}
        about DERs' curtailments $C_{p_i}$ and their weighted average. 
        
        First, average $M_i$ is invariant at $C_{p_i}=C^*$; if there are voltage violations such as $C_{max}<C^*$, $M_i$ is increasing; and if there are voltage stability margins such as $C_{min}>C^*$, $M_i$ is decreasing.
        
        Second, considering those DERs that assume the maximum (or minimum) curtailments, we have that $f_{c_i}\leq0$ (or $f_{c_i}\geq0$) for all $i \in \clN_{max}$ (or $i\in \clN_{min}$) and that $f_{c_i}<0$ (or $f_{c_i}>0$) for some $i \in \clN_{max}$ (or $i\in \clN_{min}$). This means that, in the absence of net input $u_1-u_2$, the trajectory of curtailment dynamics \eqref{eq:agent-i} is a contraction mapping that converges to the weighted mean as the consensus. In the presence of inputs $u_1$ and $u_2$, the convergence remains but to a new consensus value so long as components $u_{i1}$ and $u_{i2}$ are consistent for all $i$, respectively.    
        
        Third, inputs $u_1$ and $u_2$ have components which assume consistent values according to three distinctive cases: (a) $u_1=u_2=0$ if $C_{p_i}=C^*$ for all $i$; (b) $u_{i1}>0$ and $u_{i2}=0$ for all $i$ if there is voltage violation in the network;  (c) $u_{i1}=0$ and $u_{i2}>0$ for all $i$ if there is no voltage violation.  
        
       Combining the above three facts, we know that $C_{p_i}\rightarrow C^*$. This concludes the proof.
    \end{proof}

\bibliographystyle{IEEEtran}
\bibliography{bare_jrnl}

\begin{thebibliography}{10}
\providecommand{\url}[1]{#1}
\csname url@samestyle\endcsname
\providecommand{\newblock}{\relax}
\providecommand{\bibinfo}[2]{#2}
\providecommand{\BIBentrySTDinterwordspacing}{\spaceskip=0pt\relax}
\providecommand{\BIBentryALTinterwordstretchfactor}{4}
\providecommand{\BIBentryALTinterwordspacing}{\spaceskip=\fontdimen2\font plus
\BIBentryALTinterwordstretchfactor\fontdimen3\font minus
  \fontdimen4\font\relax}
\providecommand{\BIBforeignlanguage}[2]{{%
\expandafter\ifx\csname l@#1\endcsname\relax
\typeout{** WARNING: IEEEtran.bst: No hyphenation pattern has been}%
\typeout{** loaded for the language `#1'. Using the pattern for}%
\typeout{** the default language instead.}%
\else
\language=\csname l@#1\endcsname
\fi
#2}}
\providecommand{\BIBdecl}{\relax}
\BIBdecl

\bibitem{Alyami14}
S.~{Alyami}, Y.~{Wang}, C.~{Wang}, J.~{Zhao}, and B.~{Zhao}, ``Adaptive real
  power capping method for fair overvoltage regulation of distribution networks
  with high penetration of pv systems,'' \emph{IEEE Transactions on Smart
  Grid}, vol.~5, no.~6, pp. 2729--2738, 2014.

\bibitem{Plytaria17}
K.~E. {Antoniadou-Plytaria}, I.~N. {Kouveliotis-Lysikatos}, P.~S.
  {Georgilakis}, and N.~D. {Hatziargyriou}, ``Distributed and decentralized
  voltage control of smart distribution networks: Models, methods, and future
  research,'' \emph{IEEE Transactions on Smart Grid}, vol.~8, no.~6, pp.
  2999--3008, 2017.

\bibitem{Chai18}
Y.~{Chai}, L.~{Guo}, C.~{Wang}, Z.~{Zhao}, X.~{Du}, and J.~{Pan}, ``Network
  partition and voltage coordination control for distribution networks with
  high penetration of distributed pv units,'' \emph{IEEE Transactions on Power
  Systems}, vol.~33, no.~3, pp. 3396--3407, 2018.

\bibitem{molzahn2017survey}
D.~K. Molzahn, F.~D{\"o}rfler, H.~Sandberg, S.~H. Low, S.~Chakrabarti,
  R.~Baldick, and J.~Lavaei, ``A survey of distributed optimization and control
  algorithms for electric power systems,'' \emph{IEEE Transactions on Smart
  Grid}, vol.~8, no.~6, pp. 2941--2962, 2017.

\bibitem{xin2011cooperative}
H.~Xin, Z.~Lu, Z.~Qu, D.~Gan, and D.~Qi, ``Cooperative control strategy for
  multiple photovoltaic generators in distribution networks,'' \emph{IET
  control theory \& applications}, vol.~5, no.~14, pp. 1617--1629, 2011.

\bibitem{maknouninejad2014realizing}
A.~Maknouninejad and Z.~Qu, ``Realizing unified microgrid voltage profile and
  loss minimization: A cooperative distributed optimization and control
  approach,'' \emph{IEEE Transactions on Smart Grid}, vol.~5, no.~4, pp.
  1621--1630, 2014.

\bibitem{xin2011self}
H.~Xin, Z.~Qu, J.~Seuss, and A.~Maknouninejad, ``A self-organizing strategy for
  power flow control of photovoltaic generators in a distribution network,''
  \emph{IEEE Transactions on Power Systems}, vol.~26, no.~3, pp. 1462--1473,
  2011.

\bibitem{Xia19}
S.~{Xia}, S.~{Bu}, C.~{Wan}, X.~{Lu}, K.~W. {Chan}, and B.~{Zhou}, ``A fully
  distributed hierarchical control framework for coordinated operation of ders
  in active distribution power networks,'' \emph{IEEE Transactions on Power
  Systems}, vol.~34, no.~6, pp. 5184--5197, 2019.

\bibitem{liu2016distributed}
Y.~Liu, Z.~Qu, H.~Xin, and D.~Gan, ``Distributed real-time optimal power flow
  control in smart grid,'' \emph{IEEE Transactions on Power Systems}, vol.~32,
  no.~5, pp. 3403--3414, 2016.

\bibitem{dall2016optimal}
E.~Dall’Anese and A.~Simonetto, ``Optimal power flow pursuit,'' \emph{IEEE
  Transactions on Smart Grid}, vol.~9, no.~2, pp. 942--952, 2016.

\bibitem{dall2015photovoltaic}
E.~Dall'Anese, S.~V. Dhople, and G.~B. Giannakis, ``Photovoltaic inverter
  controllers seeking ac optimal power flow solutions,'' \emph{IEEE
  Transactions on Power Systems}, vol.~31, no.~4, pp. 2809--2823, 2015.

\bibitem{jabr2017linear}
R.~A. Jabr, ``Linear decision rules for control of reactive power by
  distributed photovoltaic generators,'' \emph{IEEE Transactions on Power
  Systems}, vol.~33, no.~2, pp. 2165--2174, 2017.

\bibitem{ecc2020}
A.~Gusrialdi, Y.~Xu, Z.~Qu, and M.~A. Simaan, ``Resilient cooperative voltage
  control for distribution network with high penetration distributed energy
  resources,'' \emph{ECC'20}, 2020, eCC'20.

\bibitem{qu2009cooperative}
Z.~Qu, \emph{Cooperative Control of Dynamical Systems}.\hskip 1em plus 0.5em
  minus 0.4em\relax Springer Science \& Business Media, 2009.

\bibitem{montenegro2015multilevel}
D.~Montenegro, G.~A. Ramos, and S.~Bacha, ``Multilevel a-diakoptics for the
  dynamic power-flow simulation of hybrid power distribution systems,''
  \emph{IEEE Transactions on Industrial Informatics}, vol.~12, no.~1, pp.
  267--276, 2015.

\bibitem{OpenSourcePlat}
R.~C. Dugan and T.~E. Mcdermott, ``An open source platform for collaborating on
  smart grid research,'' \emph{IEEE Power and Energy Society General Meeting},
  2011.

\bibitem{RATHBUN2018}
M.~Rathbun, Y.~Xu, R.~R. nejad, Z.~Qu, and W.~Sun, ``Impact studies and
  cooperative voltage control for high pv penetration,''
  \emph{IFAC-PapersOnLine}, vol.~51, no.~28, pp. 684 -- 689, 2018, 10th IFAC
  Symposium on Control of Power and Energy Systems CPES 2018.

\bibitem{Low2013branch}
M.~Farivar and S.~H. Low, ``Branch flow model: Relaxations and
  convexification—part i,'' \emph{IEEE Transactions on Power Systems},
  vol.~28, no.~3, pp. 2554--2564, 2013.

\end{thebibliography}

\begin{IEEEbiographynophoto} 
{Ying Xu}
(M’15) received the B.Eng, M.Eng, and PH.D. degrees from Harbin Institute of Technology, China, in 2003, 2005 ,and 2009 respectively. From 2009-2017, he has been a Senior Engineer in North China Grid Dispatching and Control Center. He is a postdoctoral researcher at the Department of Electrical and Computer Engineering, University of
Central Florida (UCF), USA. His main research interests and experiences include power system analysis and control, power system modeling and simulation, cooperative control, and distributed control and optimization for networked systems.
\end{IEEEbiographynophoto}


\begin{IEEEbiographynophoto}
{Zhihua Qu} (M’90-SM’93-F’09) received the Ph.D. degree in Electrical Engineering from the Georgia Institute of Technology, Atlanta, in June 1990. Since then, he has been with the University of Central Florida (UCF), Orlando. His areas of expertise are nonlinear systems and control, resilient and cooperative control, with applications to energy and power systems.
\end{IEEEbiographynophoto}
\end{document}